\documentclass{ws-ijmpa}

\begin{document}

\newlength\widd
\setlength\widd{0.48\textwidth}

\markboth{Dinko Po\v{c}ani\'c}
{A Review of Rare Pion and Muon Decays}

%
\catchline{}{}{}{}{}
%

\title{A REVIEW OF RARE PION AND MUON DECAYS}

\author{\footnotesize DINKO PO\v{C}ANI\'C}

\address{Physics Department, University of Virginia, 
         Charlottesville, VA 22904-4714, USA}

\maketitle

\pub{Received (14 July 2004)}{}

\begin{abstract}
After a decade of no measurements of pion and muon rare decays,
PIBETA, a new experimental program is producing its first results.  
We report on a new experimental study of the pion beta decay,
$\pi^+\to\pi^0 e^+\nu$, the $\pi_{e2\gamma}$ radiative decay,
$\pi^+\to e^+\nu\gamma$, and muon radiative decay, $\mu \to
e\nu\bar{\nu}\gamma$.  The new results represent four- to six-fold
improvements in precision over the previous measurements.  Excellent
agreement with Standard Model predictions is observed in all channels
except for one kinematic region of the $\pi_{e2\gamma}$ radiative
decay involving energetic photons and lower-energy positrons.

\keywords{pi mesons; muons; rare decays.}
\end{abstract}

\section{Motivation}

The PIBETA experiment\cite{pibeta} at the Paul Scherrer Institute
(PSI) is a comprehensive set of precision measurements of the rare
decays of the pion and the muon.  The goals of the experiment's first
phase are:

\begin{romanlist}[(iii)]

\item To improve the experimental precision of the pion beta decay
branching ratio, $\pi^+ \to \pi^0 e^+ \nu$ (also referred to as
$\pi_{e3}$, or $\pi_\beta$), from the present $\sim 4\,\%$ to $\sim
0.5\,\%$.

\item To measure the branching ratio of the radiative pion decay
$\pi\to e\nu\gamma$ ($\pi_{e2\gamma}$, or RPD), enabling a precise
evaluation of the pion axial-vector form factor $F_A$, and limits on
the tensor form factor $F_T$, predicted to vanish in the Standard
Model (SM).

\item An extensive measurement of the radiative muon decay rate,
$\mu\to e \nu \bar{\nu} \gamma$, with broad phase space coverage,
enabling a search for non-\,(V$-$A) admixtures in the weak Lagrangian.

\end{romanlist}

The experiment's second phase calls for a precise measurement of the
$\pi\to e \nu$ (known as $\pi_{e2}$) decay rate, used for
normalization in the first phase.  The current 0.33\,\% accuracy would
be improved to under 0.2\,\%, in order to provide a precise test of
lepton universality, and, hence, of certain extensions to the Standard
Model.  In this report we focus mainly on parts (i) and (ii) above,
and briefly discuss the first muon radiative results, (iii).

The rare pion beta decay, $\pi^+\to\pi^0e^+\nu$ (branching ratio
$R_{\pi\beta} \simeq 1 \times 10^{-8}$), is one of the most basic
semileptonic electroweak processes.  It is a pure vector transition
between two spin-zero members of an isospin triplet, and is therefore
analogous to superallowed Fermi (SF) transitions in nuclear beta
decay.  The conserved vector current (CVC)
hypothesis\cite{Ger55,Fey58} and quark-lepton universality relate the
rate of the pure vector pion beta decay to that of muon decay via the
Cabibbo-Kobayashi-Maskawa (CKM) quark mixing matrix element
$V_{ud}$\cite{Cab63,Kob73} in a theoretically exceptionally clean
way.\cite{Kal64,Sir78} Hence, pion beta decay presents an excellent
means for a precise experimental determination of the CKM matrix
element $V_{ud}$,\cite{Sir78,Sir82,Jau01,Cir02} hindered only by the
low branching ratio of the decay.

The CKM quark mixing matrix has a special significance in modern
physics as a cornerstone of a unified description of the weak
interactions of mesons, baryons and nuclei.  In a universe with three
quark generations the $3\times 3$ CKM matrix must be unitary, barring
certain classes of hitherto undiscovered processes not contained in
the Standard Model.  Thus, an accurate experimental evaluation of the
CKM matrix unitarity provides an independent check of possible
deviations from the SM.  As the best studied element of the CKM
matrix, $V_{ud}$ plays an important role in all tests of its
unitarity.  However, evaluations of $V_{ud}$ from neutron decay have,
for the most part, not been consistent with results from nuclear SF
decays.\cite{PDG04}  Clearly, a precise evaluation of $V_{ud}$ from
pion beta decay, the theoretically cleanest choice, is of interest.

Radiative pion decay offers unparalleled access to information on the
pion's structure.  Given the unique role of the pion as the
quasi-Goldstone boson of the strong interaction, the implications are
far reaching.  In the Standard Model description\cite{Bry82} of the
$\pi^+\to e^+\nu\gamma$ decay, where $\gamma$ is a real or virtual
photon ($e^+e^-$ pair), the decay amplitude ${\cal M}$ depends on the
vector V and axial vector A weak hadronic currents. Both currents give
rise to structure-dependent terms SD$_{\rm V}$ and SD$_{\rm A}$
associated with virtual hadronic states, while the axial-vector
current alone causes the inner bremsstrahlung process IB from the pion
and positron.  The IB contribution to the decay probability can be
calculated in a straightforward manner using QED methods. The
structure-dependent amplitude is parameterized by the vector form
factor $F_V$ [constrained by CVC to 0.0259(5)] and the axial vector
form factor $F_A$ that have to be extracted from experiments.

The ratio $\gamma = F_A/F_V$ in $\pi\to e\nu\gamma$ decay directly
determines the chiral perturbation theory parameter sum
$(l_9+l_{10})$, or, equivalently, $\alpha_E$, the pion
polarizability.\cite{Hol90,Bij97,Gen03} These quantities are of
longstanding interest since they are among the few unambiguous
predictions of chiral symmetry and QCD at low energies.  The current
status of these measurements is not satisfactory, as there is
considerable scatter among the various experimental determinations of
$F_A/F_V$, and the accepted Particle Data Group (PDG) average value
has a 14\,\% uncertainty.\cite{PDG04}

Moreover, the statistical accuracy of the present experimental data on
the radiative pion decay\cite{Dep63p,Ste74,Ste78,Bay86,Pii86,Bol90}
cannot rule out contributions from other allowed terms in the
interaction lagrangian, namely the scalar S, pseudoscalar P, and
tensor T admixtures.\cite{Mur85} Nonzero values of any of these
amplitudes would imply new physics outside the Standard Model.  In
particular, reports from the ISTRA collaboration\cite{Pob90,Pob03}
have indicated a nonzero tensor term, $F_T = -0.0056\,(17)$.  A
careful analysis by Herczeg of the existing beta decay data set could
not rule out such a value of $F_T$, which he presumed could be due to
leptoquarks\cite{Her94}.  In contrast, Chizhov proposed a new
intermediate chiral boson with an anomalous interaction with matter,
in order to account for the apparent non-(V$-$A) behavior in
RPD.\cite{Chi93}

Finally, radiative muon decay (RMD) provides an even better testing
ground for non-(V$-$A) interaction terms than does RPD, thanks to the
absence of internal structure in the muon.  Measuring photon--positron
energy distributions in RMD allows an evaluation of the $\mu^+$ decay
parameter $\bar{\eta}$ that is predicted to be zero in the V$-$A
Standard Model.  The current limit on $\bar{\eta}$ is loose, $0.02 \pm
0.08$.  It is of considerable interest to bring the uncertainties down
to a level competitive with the constraints from standard muon decay.

\begin{figure}[b]
\hbox to \textwidth{\hfill
\resizebox{0.7\textwidth}{!}{\includegraphics{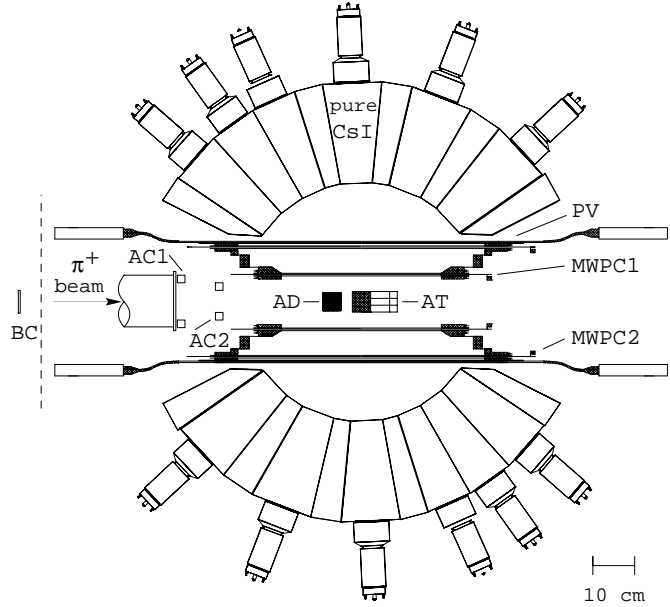}}
                    \hfill}
\vspace*{8pt}
\caption{Schematic cross section of the PIBETA apparatus
   showing the main components: beam entry counters (BC, AC1, AC2),
   active degrader (AD), active target (AT), wire chambers (MWPCs) and
   support, plastic veto (PV) detectors and PMTs, pure CsI calorimeter
   and PMTs.}
\label{fig:det:xsect}
\end{figure}

\section{Experimental Method}

The PIBETA apparatus is a large-acceptance non-magnetic detector
optimized for detection of photons and electrons in the energy range
of 5--150$\,$MeV with high efficiency, energy resolution and solid
angle.  The main sensitive components of the apparatus, shown and
labeled in Fig.~\ref{fig:det:xsect}, are: 
\begin{romanlist}[(iii)] 

\item beam defining plastic scintillator detectors: BC, a thin
forward beam counter, AC$_1$ and AC$_2$, cylindrical active
collimators, AD, an active degrader, AT, a 9-element segmented active
target that stops the $\pi^+$ beam;
\item charged particle tracking and id: MWPC$_1$ and MWPC$_2$,
cylindrical chambers, and PV, a 20-bar segmented thin plastic
scintillator hodoscope;
\item a 240-element segmented spherical pure-CsI shower
calorimeter, subtending a solid angle of $\sim 80\,$\% of $4\pi$.
\end{romanlist}
The entire system is enclosed in a temperature-controlled Pb house
lined with cosmic muon veto detectors.

To collect the available $\pi_beta$ decay events we recorded all
non-prompt large-energy (above the $\mu \to e\nu\bar{\nu}$ endpoint)
electromagnetic shower pairs occurring in opposite detector
hemispheres (non-prompt two-arm events).  In addition, we recorded a
large prescaled sample of non-prompt single shower (one-arm) events.
Using these minimum-bias sets, we extract $\pi_\beta$ and $\pi_{e2}$
events, using the latter for normalization.  In a stopped pion
experiment these two channels have nearly the same detector
acceptance, and have much of the systematics in common.

These two event classes are part a full complement of twelve fast
analog triggers comprising all relevant logic combinations of one- or
two-arm, low- or high calorimeter threshold , prompt and delayed (with
respect to $\pi^+$ stop time), as well as a random and a three-arm
trigger, all of which were implemented in order to obtain maximally
comprehensive and unbiased data samples.  Details of the method are
explained in detail in Ref.~\refcite{Frl04}.  The list of decays
measured and/or used for normalization in our work is given in
Table~\ref{tab:decays}.

\begin{table}[h]

\tbl{List of decays measured in the PIBETA experiment, along with
the corresponding branching ratios.\label{tab:decays}}
{\begin{tabular}{llcl} \toprule

\multicolumn{2}{c}{Decay} & Branching ratio \\
\colrule
$\pi^+\to$ & $\mu^+ \nu$ & 1.0                               \\
    & $\mu^+\nu\gamma$ & $\sim 2.0 \times 10^{-4}$           \\
    & e$^+ \nu$   & $\sim 1.2 \times 10^{-4}$ & normalize to \\
    & e$^+\nu\gamma$ & $\sim 5.6 \times 10^{-8}$ & measure   \\
    & $\pi^0$e$^+\nu$ & $\sim 1.0 \times 10^{-8}$ & measure  \\[1ex]

$\pi^0\to$ & $\gamma\gamma$ & $\sim 0.9880$ &  measure       \\
    & e$^+$e$^-\gamma$ & $\sim 1.2 \times 10^{-2}$ & measure \\
    & e$^+$e$^-$e$^+$e$^-$ & $\sim 3.1 \times 10^{-5}$       \\
    & e$^+$e$^-$ & $\sim 6.2 \times 10^{-8}$ &               \\[1ex]

$\mu^+\to$ & e$^+\nu\overline{\nu}$ & 1.0 &  normalize to    \\
    & e$^+\nu\overline{\nu}\gamma$ & $\sim 0.014$ &  measure \\
    & e$^+\nu\overline{\nu}$e$^+$e$^-$ 
                        & $\sim 3 \times 10^{-5}$ &  measure \\ 
\botrule
\end{tabular}}
\end{table}

\section{Results}

The building and testing of the detector components were completed in
1998, followed by the assembly and commissioning of the full detector
apparatus.  Data acquisition with the PIBETA detector started in the
second half of 1999, initially at a reduced pion stopping rate, as
planned.  The experiment ran subsequently during 2000 and 2001 at
$\sim 1\,$MHz $\pi^+$ stopping rate.  We review below the current
status of the analysis of the data acquired during this running
period.

\subsection{Pion beta decay, $\pi^+\to\pi^0e^+\nu$}

\begin{figure}[h]
\noindent\hbox to\textwidth{
\resizebox{\widd}{!}{\includegraphics{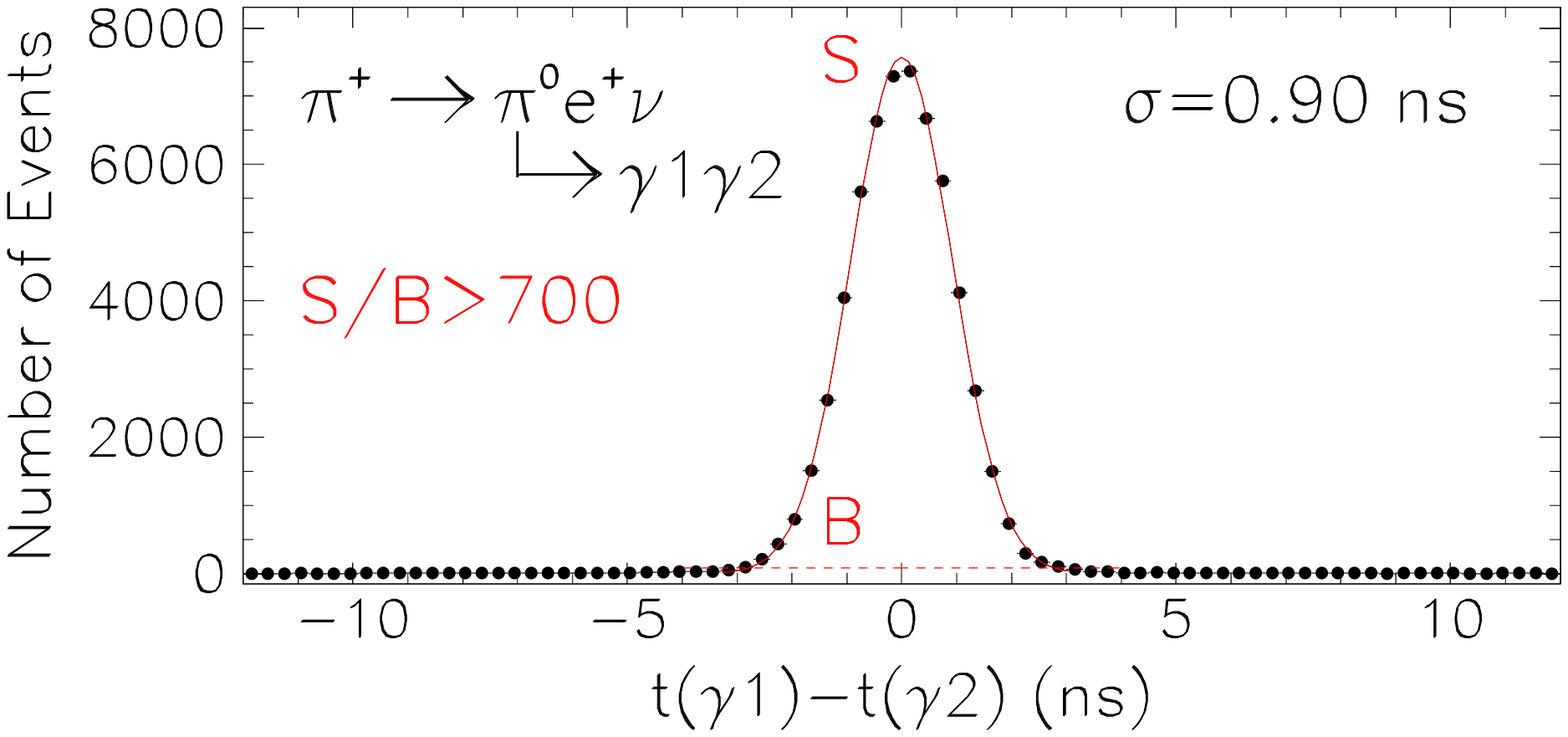}}  \hfil
\resizebox{\widd}{!}{\includegraphics{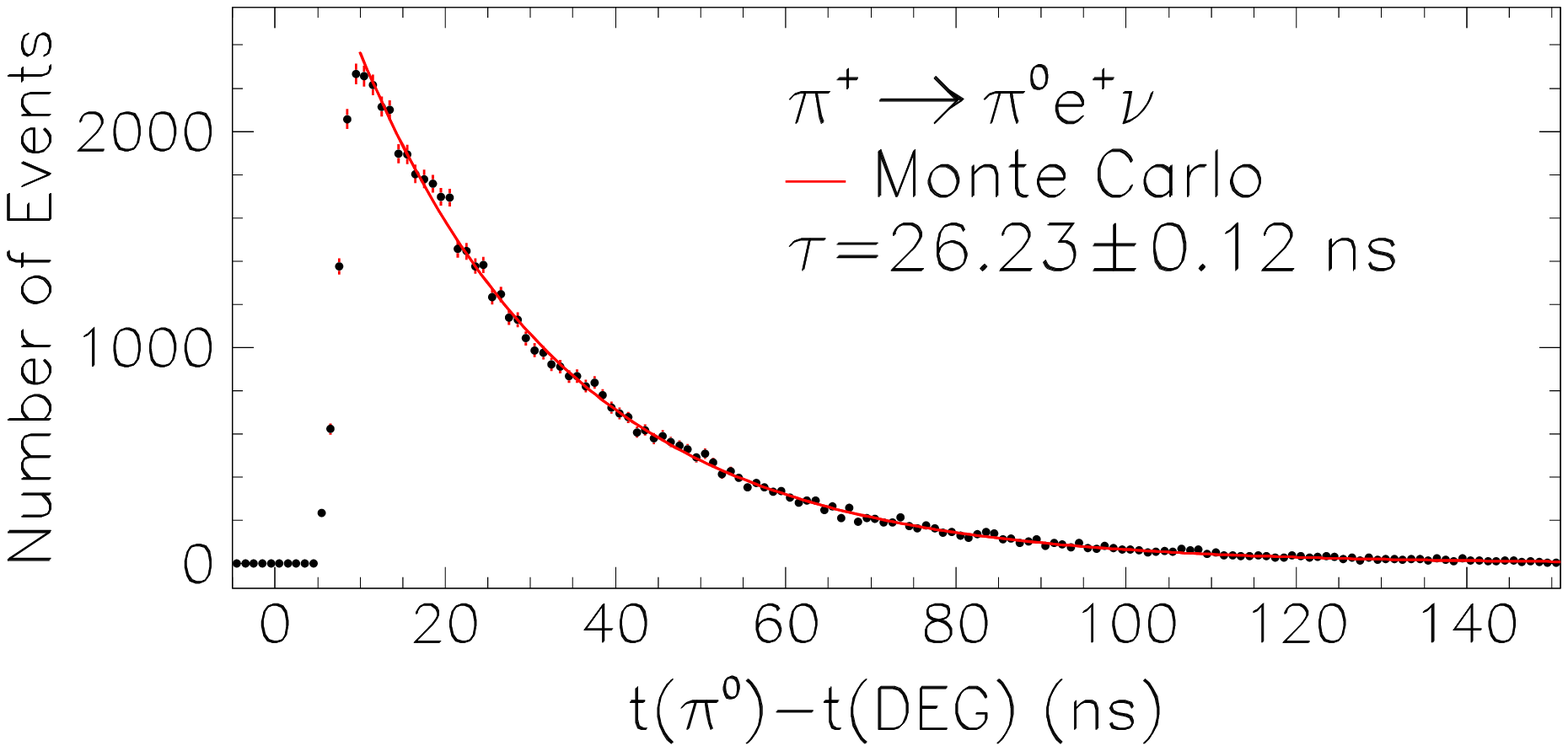}}
                           }
\noindent\hbox to\textwidth{
\resizebox{\widd}{!}{\includegraphics{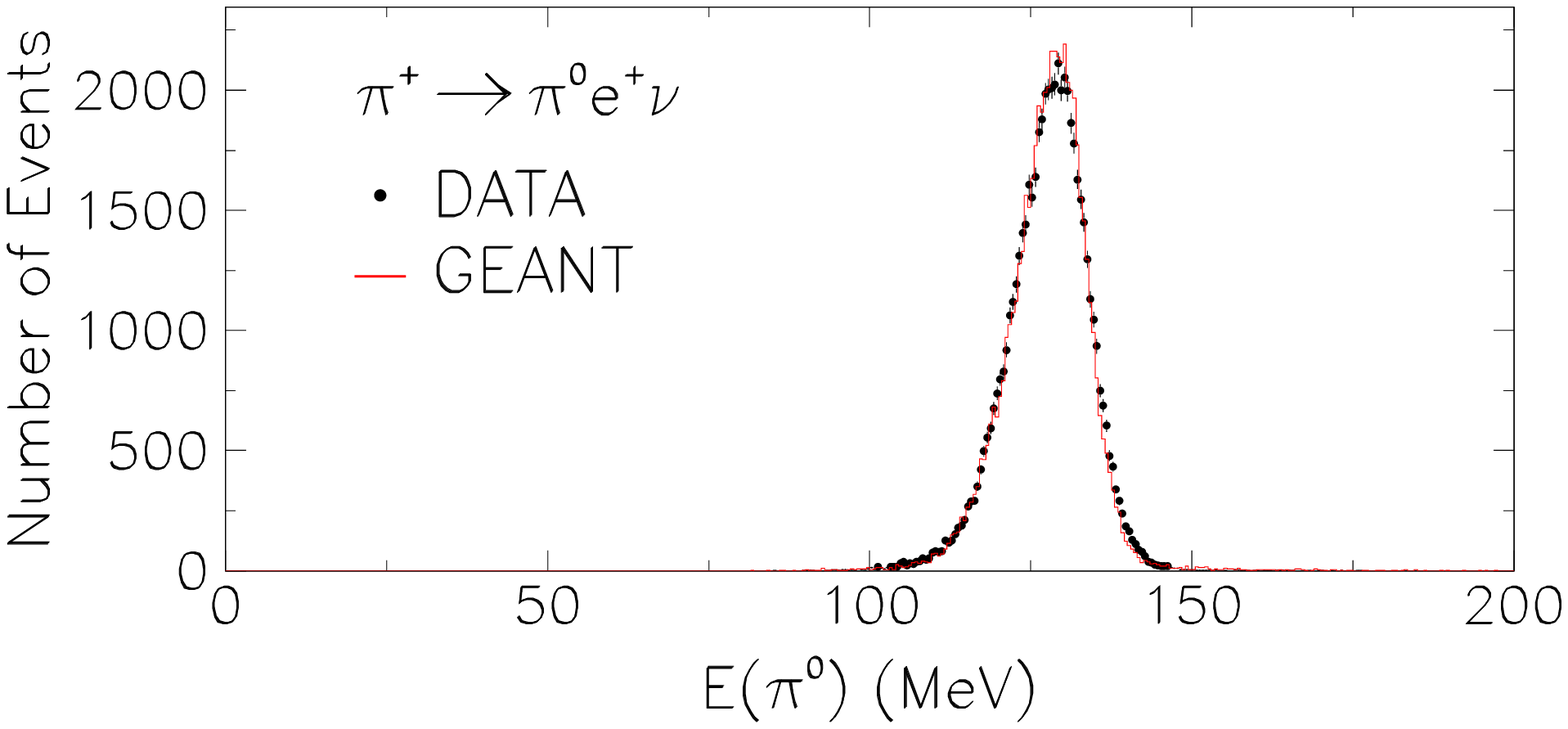}}  \hfil
\resizebox{\widd}{!}{\includegraphics{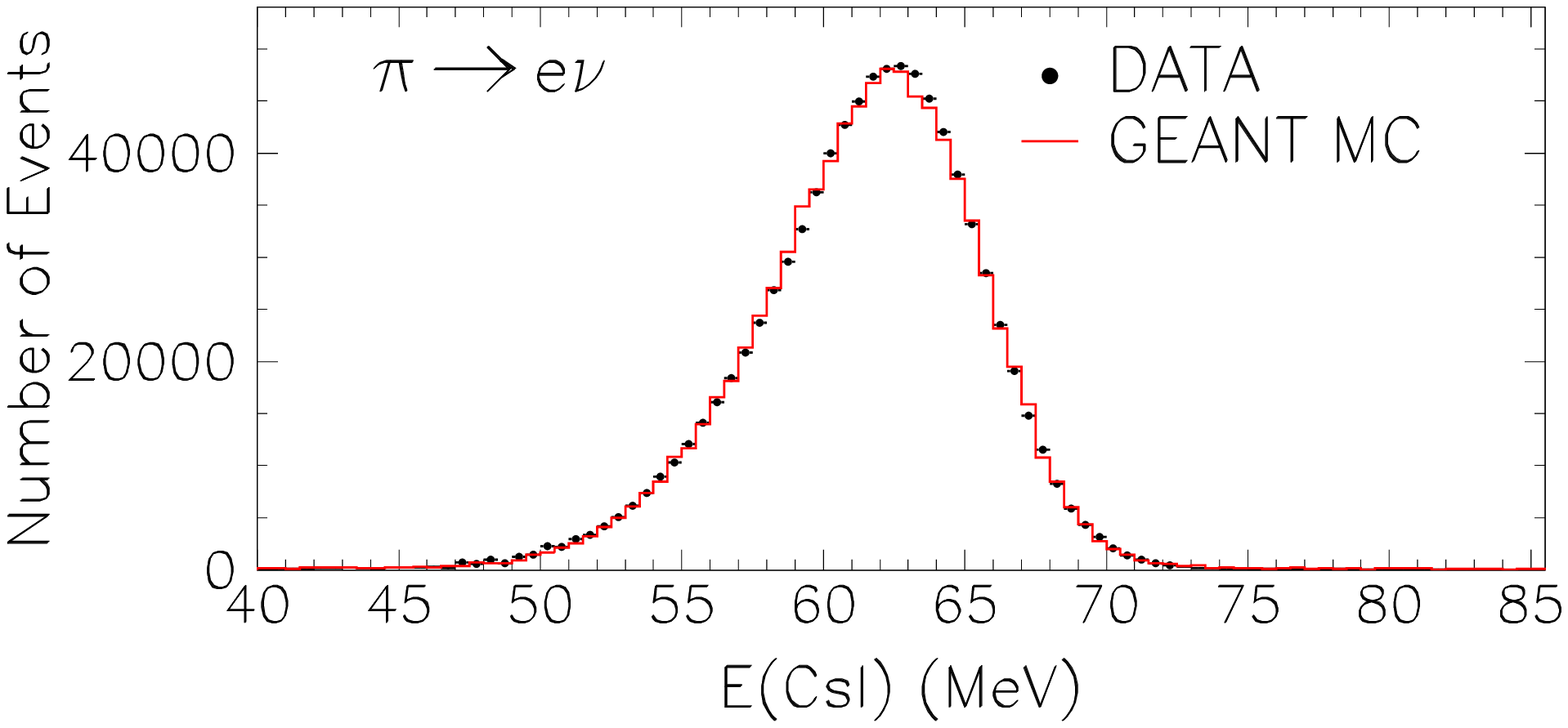}}
                           }
\caption{Histogram of $\gamma$-$\gamma$ time differences for
$\pi_\beta$ decay events (dots); curve: fit with a Gaussian function
plus a constant (top left).  Histogram of time differences between the
beam pion stop and the $\pi_\beta$ decay events (dots); curve: pion
lifetime (top right).  CsI calorimeter energy spectra for the pion
beta decay events (bottom left) and the normalizing decay $\pi_{2e}$
(bottom right).}
\label{fig:pb:sn:tim}
\end{figure}

In spite of its low branching ratio, the pion beta decay signal,
marked by a nearly back-to-back energetic pair of neutral showers from
the $\pi^0 \to\gamma\gamma$ decay, is strong and without appreciable
background.  This is demonstrated in Fig.~\ref{fig:pb:sn:tim} which
shows virtually background-free histograms of $\gamma$-$\gamma$
relative timing, and of the event time distribution following the pion
stop time, $t$({\sc deg}).  The corresponding CsI calorimeter energy
distributions for the $\pi_\beta$ events and for the normalizing
$\pi_{e2}$ events are also  given in the same figure 
We note the
excellent agreement between the measured distributions and those
simulated by the Monte Carlo program GEANT3.

Details of the analysis method of the $\pi_\beta$ decay channel can be
found in Refs.~\refcite{Poc03} and \refcite{Wli04}.  Applying our
method and using the PDG 2004 recommended value of $R_{\pi e2}^{\rm
exp}=1.230(4)\times 10^{-4}$,~\cite{PDG04} we extract the pion beta
decay branching ratio:
\begin{equation}
   R^{\rm exp}_{\pi\beta} = \rm 
      [1.036 \pm 0.004\,(stat) \pm 0.004\,(syst) \pm 0.003\,(\pi_{e2})]
                   \times 10^{-8}\,,   \label{eq:pb_br}
\end{equation}
or, in terms of the decay rate,
\begin{equation}
   \Gamma^{\rm exp}_{\pi\beta} = \rm
  [0.3980 \pm 0.0015(stat) \pm 0.0015(syst) 
                                      \pm 0.0013(\pi_{e2})]\,s^{-1}\,.
\end{equation}
In both expressions the first uncertainty is statistical, the second
systematic, and the third is the $\pi_{e2}$ branching ratio
uncertainty.  Our result represents a six-fold improvement in accuracy
over the most precise previous measurement\cite{McF85}.  Furthermore,
our result is in excellent agreement with predictions of the SM and
CVC given the PDG recommended value range for $V_{ud}$:\cite{PDG04}
\begin{equation}
    R^{{\rm SM}}_{\pi\beta} 
          = (1.038 - 1.041) \times 10^{-8} \quad {\rm (90\% C.L.)}\,,
                                    \label{eq:pb_sm_pred}
\end{equation}
and represents the most accurate test of CVC and Cabibbo universality
in a meson to date.  Our result confirms the validity of the radiative
corrections for the process at the level of $4\sigma_{{\rm exp}}$,
since, excluding loop corrections, the SM would predict
$R^{\rm no\ rad.\ corr.}_{\pi\beta} = (1.005 - 1.007) \times
10^{-8}$ at 90\,\% C.L. 

Using our measured branching ratio $R^{{\rm exp}}_{\pi\beta}$, we can
calculate a new value of $V_{ud}$ from pion beta decay,
$V_{ud}^{\rm (PIBETA)} = 0.9728(30)$, which is in excellent
agreement with the PDG 2004 average, $V_{ud}^{\rm (PDG'04)} =
0.9738(5)$.  We will continue to improve the overall accuracy of the
$\pi\beta$ decay branching ratio to $\sim$\,0.5\,\% by further
refining the experiment simulation and analysis, and by adding new
data.

\subsection{Radiative pion decay, $\pi^+ \to e^+\nu\gamma$}

During our 1999-2001 run we recorded over 40,000 $\pi_{e2\gamma}$
radiative pion decays (RPD) events.  Besides its intrinsic interest,
the $\pi_{e2\gamma}$ process is an important physics background to
other decays under study, particularly the $\pi_\beta$ decay.  The
pion radiative decay analysis has given us the most surprising result
to date, and has commanded significant effort on our part to resolve
the issue.

The different event triggers used in our experiment are sensitive to
three distinct regions in the $\pi_{e2\gamma}$ phase space:

\begin{itemlist}

\item region A with $e^+$ and $\gamma$ emitted into opposite
hemispheres, each with energy exceeding that of the Michel edge ($E_M
\simeq 52\,$MeV), recorded in the main two-arm trigger,

\item region B with an energetic photon ($E_\gamma > E_M$), and
$E_{e+} \geqslant 20\,$MeV, recorded in the one-arm trigger, and

\item region C with an energetic positron ($E_{e+} > E_M$), and
$E_\gamma \geqslant 20\,$MeV, also recorded in the one-arm trigger.

\end{itemlist}

\begin{figure}
\noindent\hbox to\textwidth{
\resizebox{\widd}{!}{\includegraphics{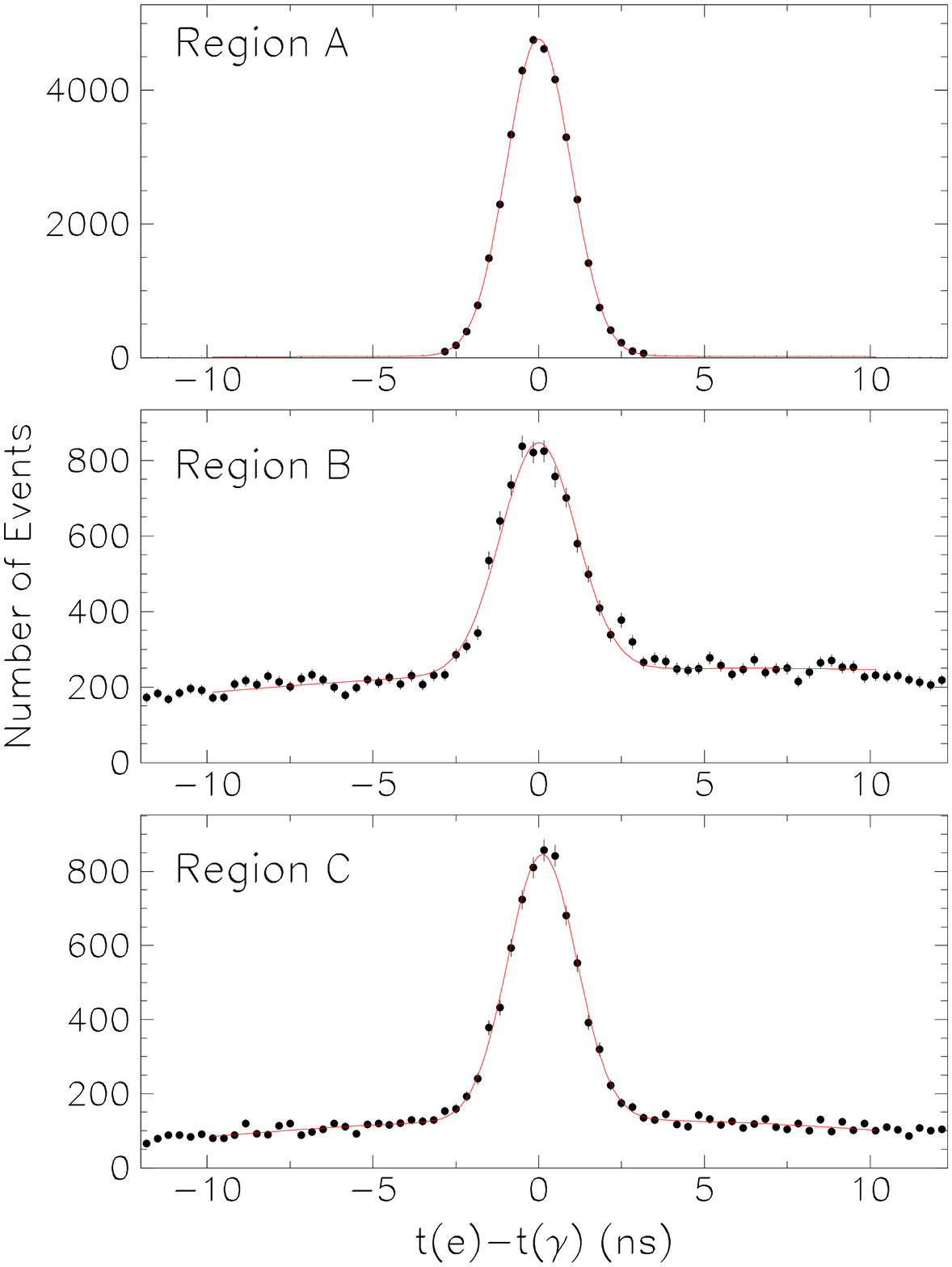}}  \hfil
\resizebox{\widd}{!}{\includegraphics{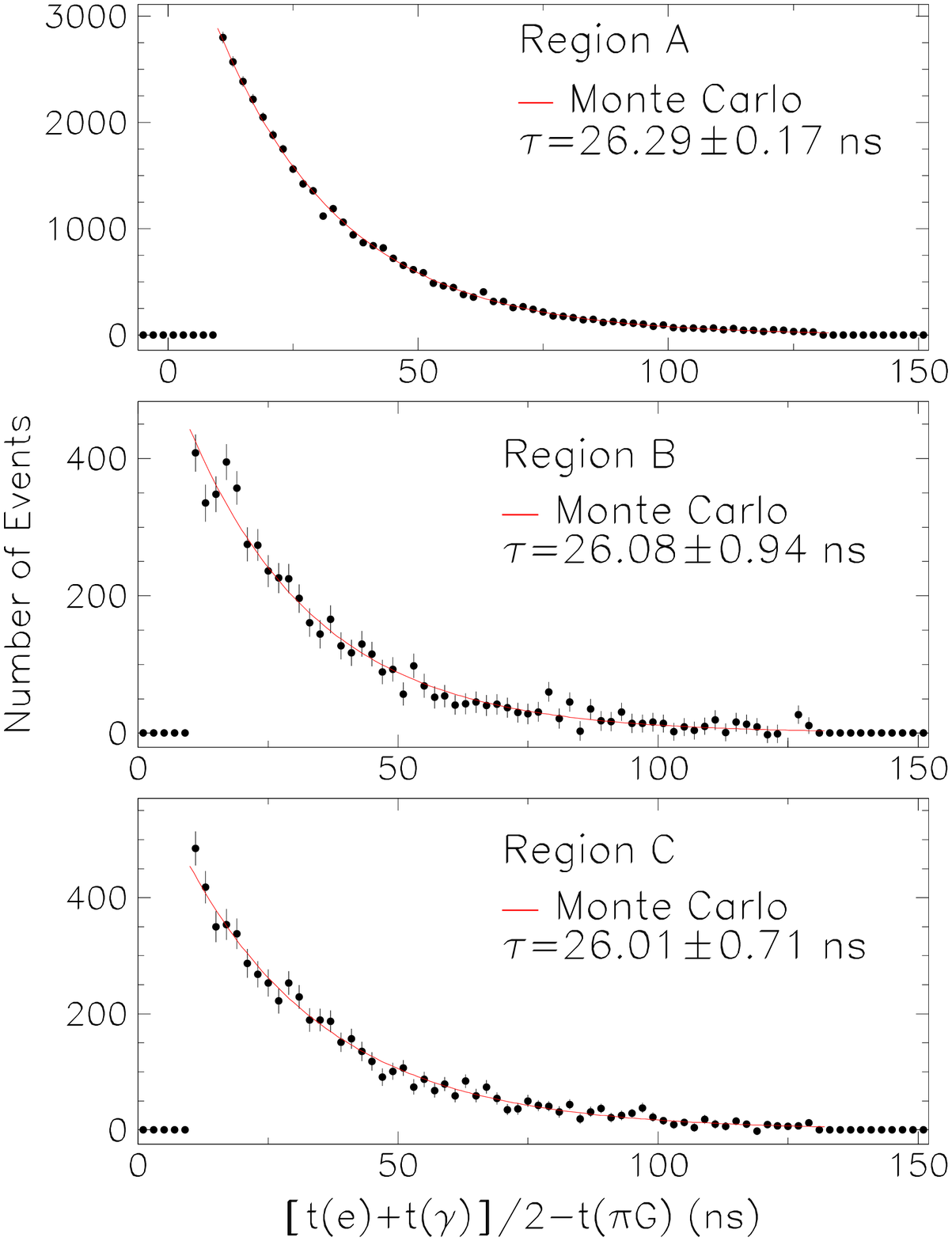}}
                           }
\caption{$e$-$\gamma$ timing difference for $\pi\to e\nu\gamma$ decay
events in regions A, B, and C (left panels, top to bottom,
respectively).  The right panels plot the $\pi^+\to e^+\nu\gamma$
event timing relative to the $\pi^+$ stop gate time, $t(\pi{\rm G})$,
after accidental background subtraction.  Monte Carlo decay functions
are shown as full lines; best-fit values for the pion lifetime are
indicated for each region.}
\label{fig:pe2g:sn:tim}

\end{figure}

Together, the three regions overconstrain the Standard Model
parameters describing the decay, and thus allow us to examine possible
new information about the pion's hadronic structure, or non-(V$-$A)
interactions.  The RPD data are of a similar quality to our
$\pi_\beta$ event set, particularly in region A, as is readily
verified in Fig.~\ref{fig:pe2g:sn:tim}, which shows histograms of
$e$-$\gamma$ time differences and event timing with respect to the
pion stop gate time, $t(\pi\rm G)$.  The analysis of these data is
involved; more details can be found in Refs.~\refcite{Frl03b} and
\refcite{pibeta}.

\begin{figure}[t]
\hbox to \textwidth{
\includegraphics[width=\widd]{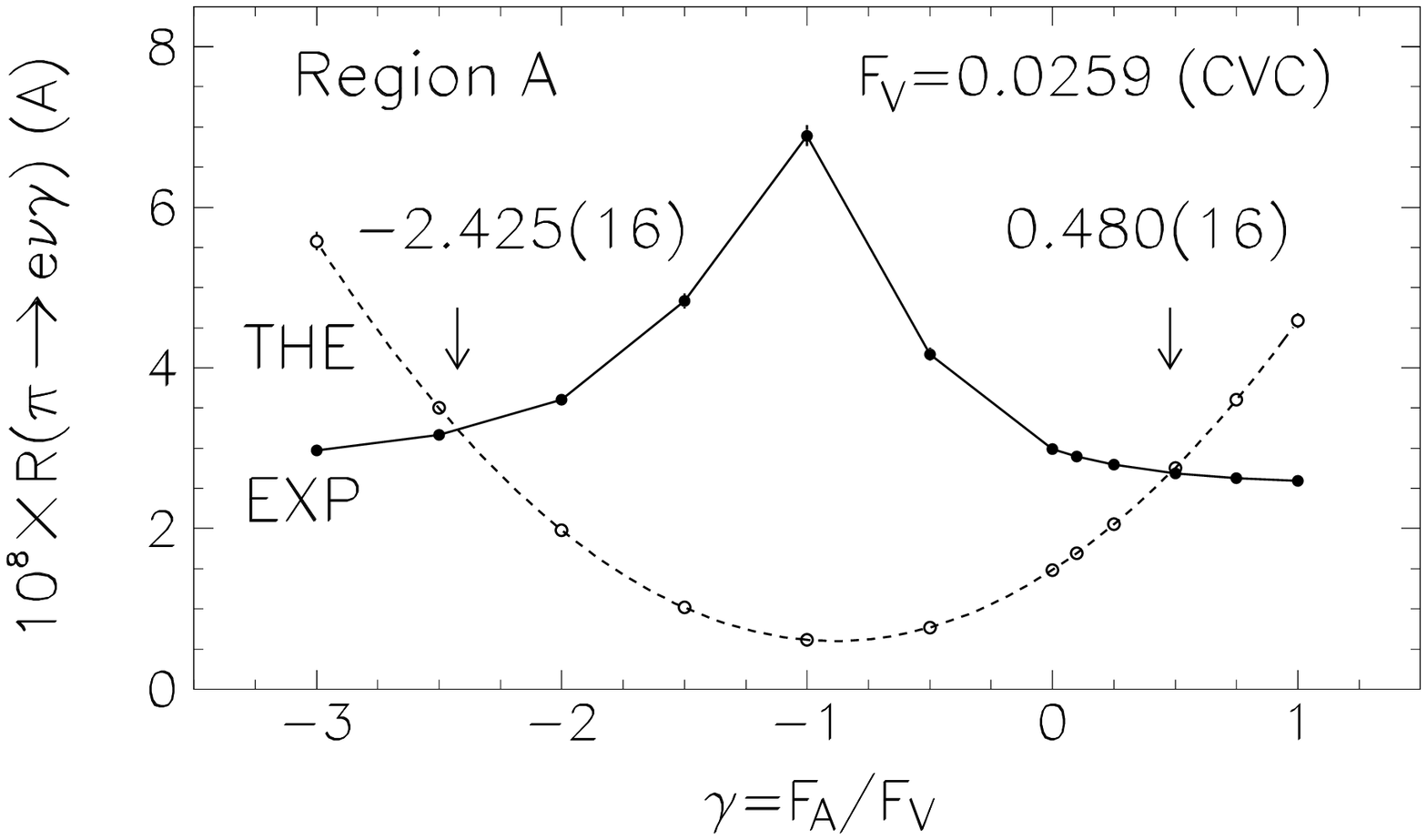} \hfil
\includegraphics[width=\widd]{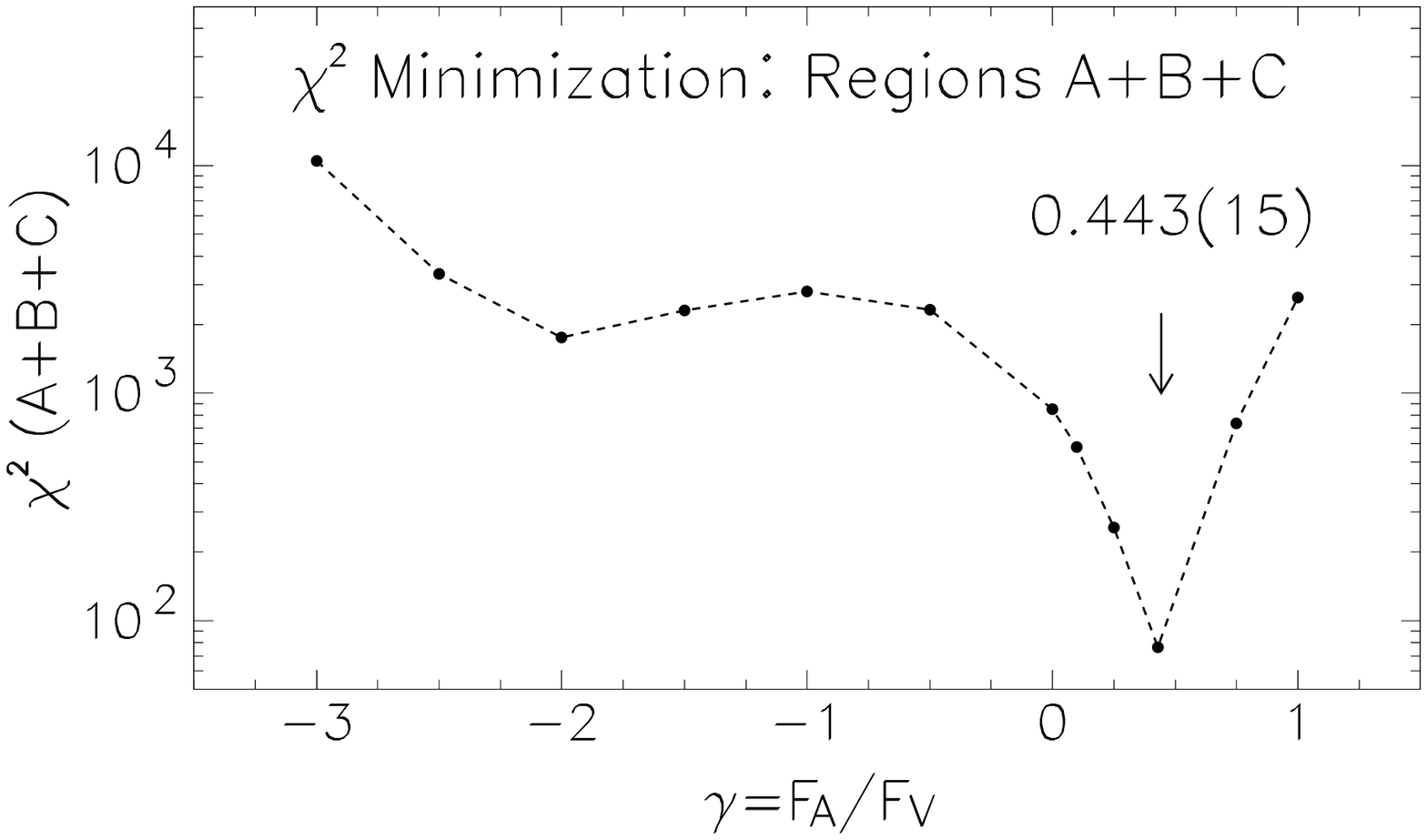}}
\caption{Left plot: $\pi^+\to e^+\nu\gamma$ branching ratio values as
a function of $\gamma\equiv F_A/F_V$.  The theoretical parabola
follows from the $V - A$ model).  The experimental values reflect fits
to region $A$ data only.  Right plot: minimum $\chi^2$ values of
simultaneous fits to the entire data set (regions $A$, $B$, $C$).}
\label{fig:rp_br}
\end{figure}

The dependence of the region-$A$ experimental and theoretical
branching ratio on the value of $\gamma$ is shown in
Fig.~\ref{fig:rp_br} (left), indicating two solutions.  The positive
$\gamma$ solution is preferred by a $\chi^2$ ratio of $\sim\,50$:1
once data from regions $B$ and $C$ are included in the analysis
(right panel).  We compare the experimental and theoretical branching
ratios for the three phase space regions in Table~\ref{tab:p2eg:br}.
We note that due to the large statistical and systematic uncertainties
present in all older experiments, our values are consistent with
previously published measurements.  The best CVC fit to our data
yields 
\begin{equation}
    \gamma = 0.443\pm 0.015\,, \quad {\rm or} \qquad
       F_A=0.0115(4) \quad {\rm with} \quad F_V \equiv 0.0259\,.
\end{equation}
This result represents a four-fold improvement in precision
over the previous world average $F_A=0.0116(16)$ \cite{PDG04}.  It is
consistent with chiral Lagrangian
calculations~\cite{Hol86,Bij97,Gen03}, and will lead to a
correspondingly improved precision in the order $p^4$ chiral constant
$l_{10}$ \cite{Amo01,Gen03}.
 
\begin{table}
\tbl{Best-fit $\pi\to e\nu\gamma$ branching ratios.
\label{tab:p2eg:br}}
{\begin{tabular}{ccccc} \toprule
 $E^{\rm min}_{e^+}$ & $E^{\rm min}_\gamma$ 
                 & $\theta^{\rm min}_{e\gamma}$  
                                & $R_{\rm exp}$ & $R_{\rm the}$ \\
 (MeV) & (MeV) &       & $(\times 10^{-8})$ & $(\times 10^{-8})$  \\
\colrule
 $50$  & $50$  & $-$        & $2.71(5)$ & $2.583(1)$ \\
 $10$  & $50$  & $40^\circ$ & $11.6(3)$ & $14.34(1)$ \\
 $50$  & $10$  & $40^\circ$ & $39.1(13)$ & $37.83(1)$  \\
\botrule
\end{tabular}}
\end{table}

\begin{figure} [b]
\hbox to \textwidth{
\includegraphics[width=\widd]{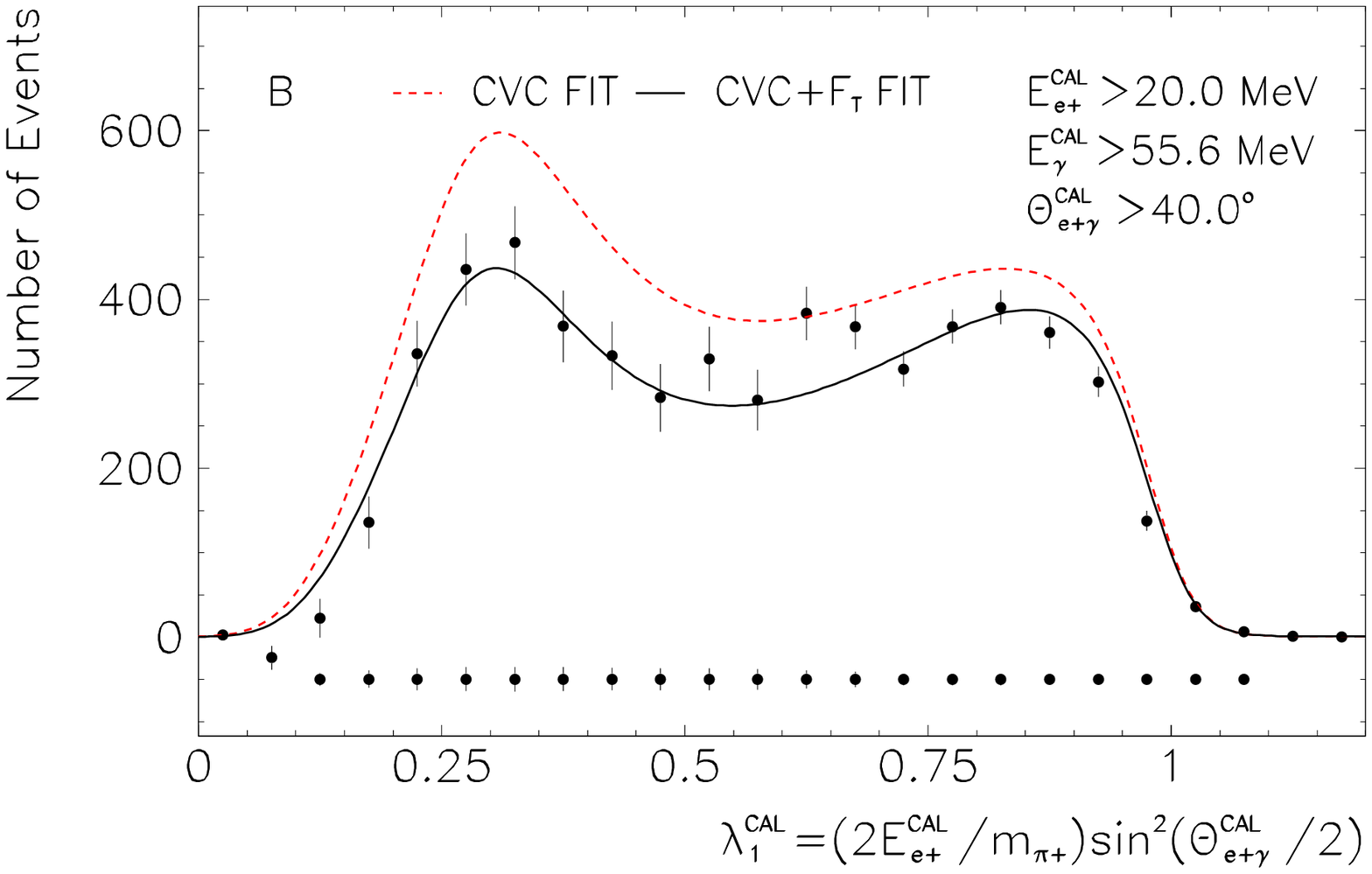} \hfil
\includegraphics[width=\widd]{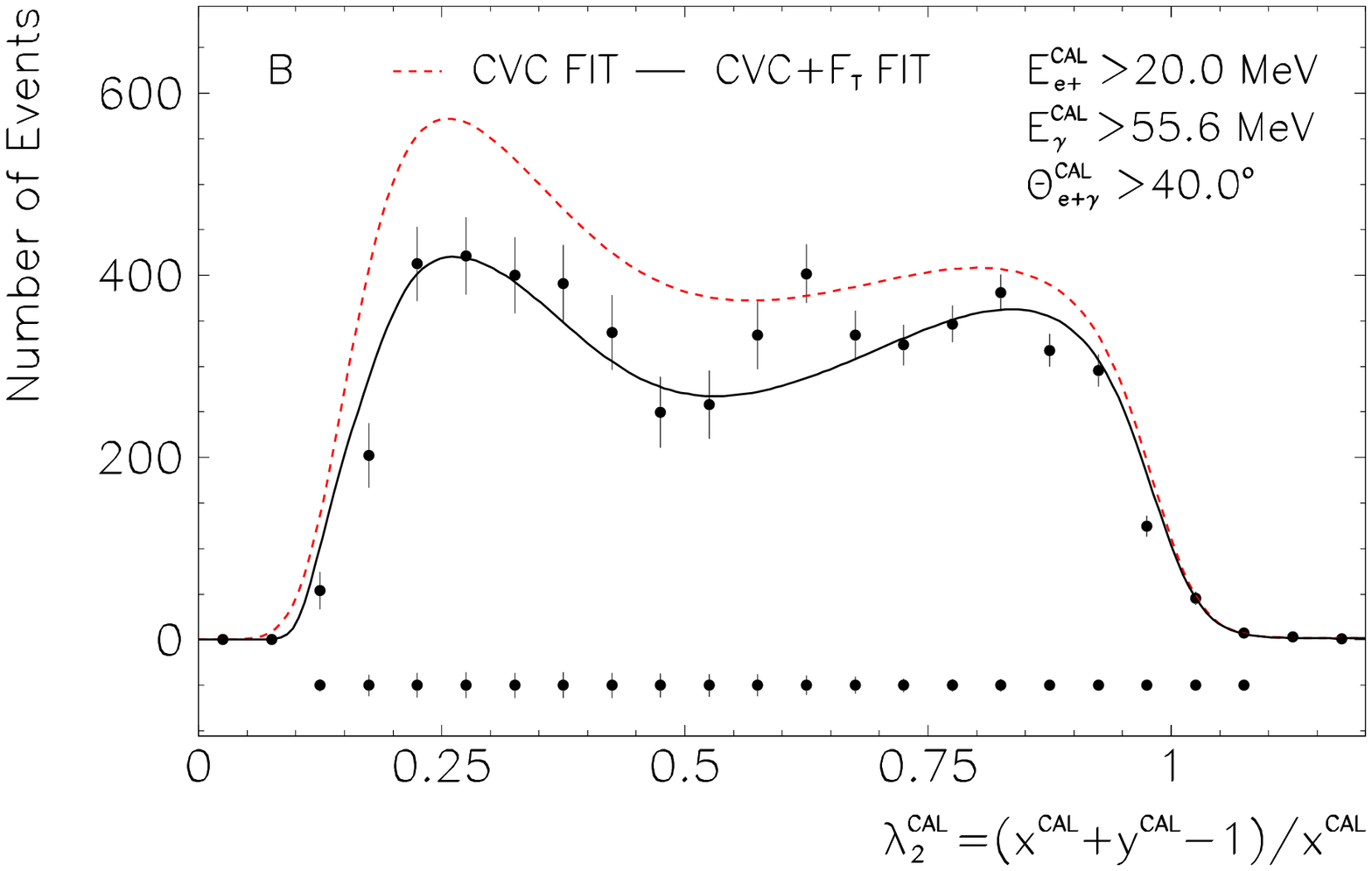}}
\caption{Left panel: measured spectrum of $\lambda_1 = (2E_e/m_\pi)
\sin^2(\theta_{e\gamma}/2)$ in $\pi_{e2\gamma}$ decay for the
kinematic region $B$, with limits noted in the figure.  Dashed curve:
three-region global best fit with the pion form factor $F_V = 0.0259$
fixed by the CVC hypothesis, $F_T=0$, and $F_A$ free.  Solid curve:
$F_V = 0.0259$ and $F_A = 0.0115$ from the first fit, this time with
$F_T$ released to vary freely, resulting in $F_T = -0.0018\,(3)$.
Error bars on the points at bottom of graph reflect the expected
uncertainties in the proposed dedicated measurement.
Bottom panel: same as above, but plotting the variable $\lambda$
evaluated purely on the basis of photon and positron energies:
$\lambda_2 = (x+y-1)/x$, where $x=2E_\gamma/m_\pi$, and
$y=2E_{e}/m_\pi$.  The two methods agree well. 
\label{fig:rp2a_l12prop}}
\end{figure}

Thus, our experimental $\pi^+\to e^+\nu\gamma$ branching ratios and
energy distributions in kinematic regions $A$ and $C$ are compatible
with the ($V-A$) interaction.  The sizable 19\,\% shortfall in the
measured branching ratio in region $B$ dominates the high value of
$\chi^2/{\rm d.o.f} = 25.4$ (Table~\ref{tab:p2eg:br}), and is
disconcerting.  In a fit restricted to region $A$ data only, we obtain
$\gamma = 0.480\pm 0.016$; this result remains unchanged if region $C$
data are added to the fit.  Significantly, all previous studies except
one\cite{Bol90} (which, too, found an anomaly), have analyzed only
data with kinematics compatible with our region $A$.  An illustration
of the nature of the observed discrepancy is given in the region-$B$
plots of a conveniently defined kinematic variable $\lambda$ in
Fig.~\ref{fig:rp2a_l12prop}.  Addition of a small negative tensor
term, $F_T \sim -0.002$, improves agreement with the data.

As of this writing the PIBETA collaboration is pursuing a dedicated
run to determine the extent and nature of the observed discrepancy
more precisely.

\subsection{Radiative muon decay, $\mu^+ \to e^+\nu\bar{\nu}\gamma$}

\begin{figure}[h]
\hbox to \textwidth{
 \resizebox{\widd}{!}{\includegraphics{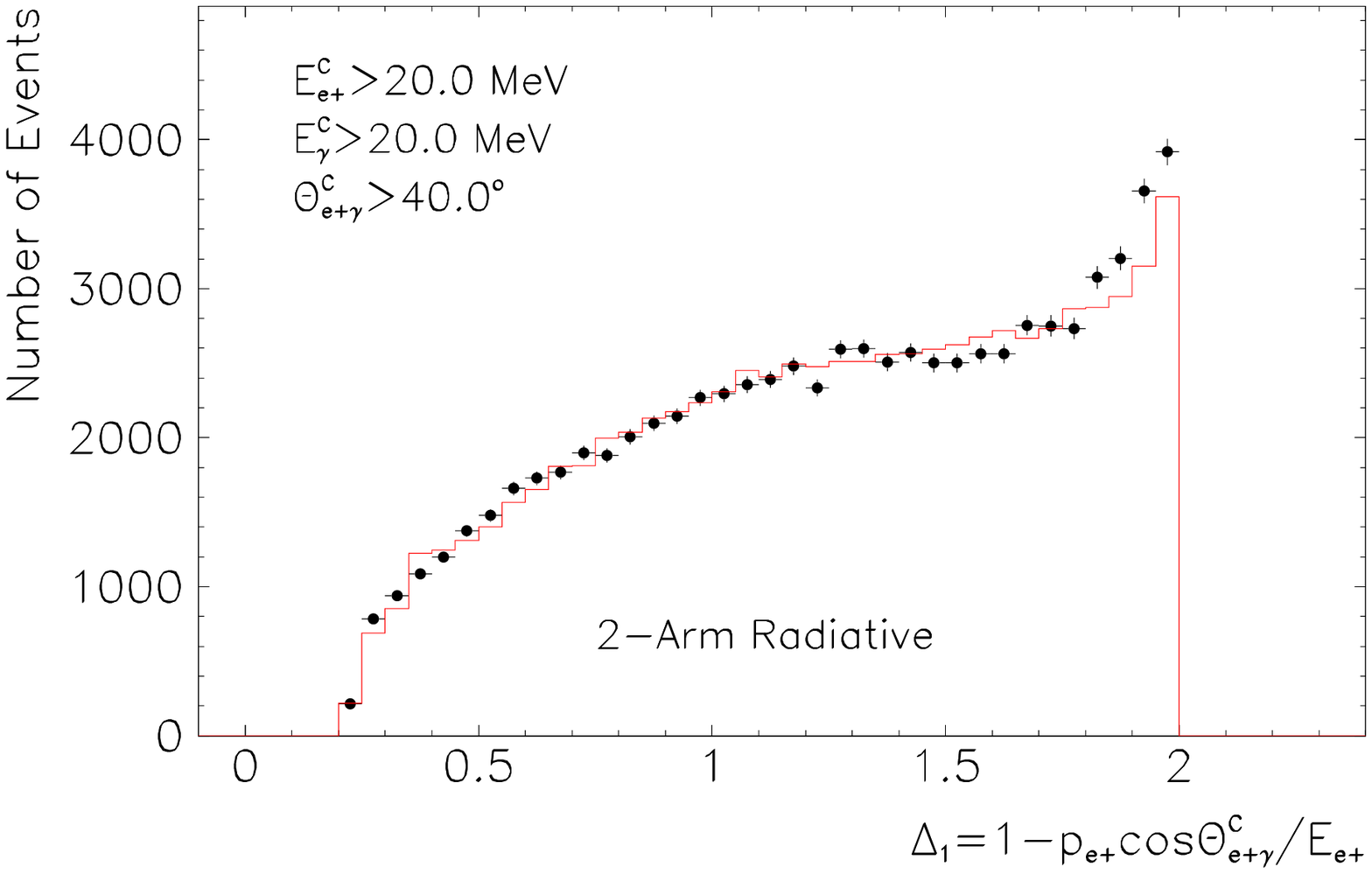}} \hfil
 \resizebox{\widd}{!}{\includegraphics{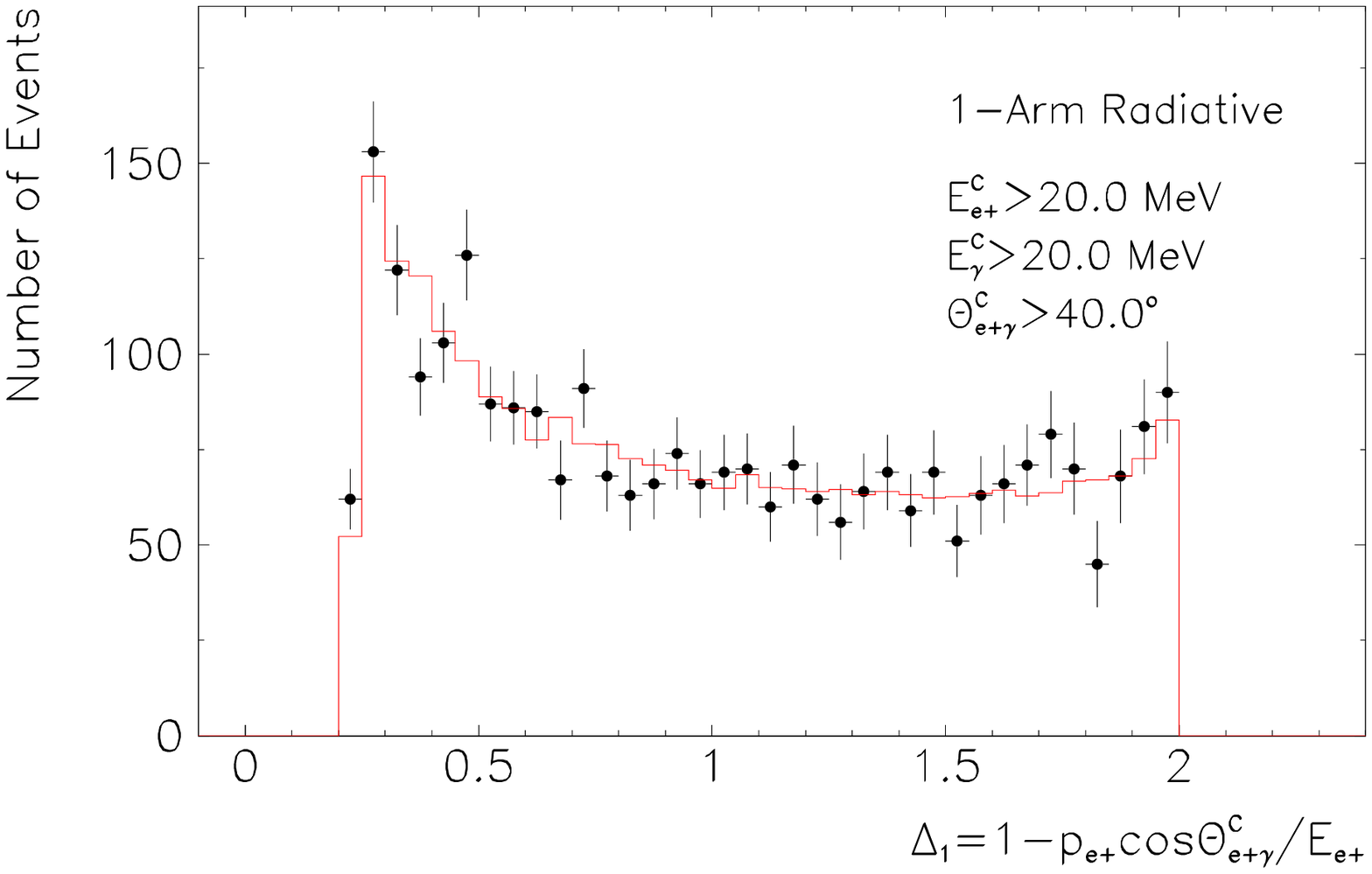}}
                   }
\caption{Dots: distributions of the kinematic variable $\Delta = 1 -
p_e\cos\theta_{e\gamma}/E_e$ for radiative muon decay events collected
with the two-arm (left) and one-arm trigger (right).  Histograms:
Standard Model V$-$A GEANT simulation.}
\label{fig:rmd:delta}
\end{figure}

The 1999-2001 run produced a set of some 300,000 radiative muon decay
events, increasing the world set by two orders of magnitude.  
Due to an emphasis in the early analysis on the pion rare
decay channels, 
the RMD analysis has not yet reached the same sub-1\,\% level of
precision.  The analysis is currently at the $\sim 1$\,\% level and
both integrated and differential branching ratios are in good
agreement with the V$-$A SM predictions at this level.\cite{Frl03a}
Of particular interest are the shapes of differential distributions of
$\Delta$, a suitably defined kinematic variable (analogous to $\lambda$
in RPD), shown in Fig.~\ref{fig:rmd:delta}.  Clearly, nothing like the
$\sim 20$\,\% discrepancy of region $B$ in RPD is observed here.

This work is in progress, and the 2004 dedicated RPD run is expected
to more than double the RMD event set, adding high-quality new data.
Interpretation of the PIBETA RMD data below the 1\,\% accuracy level
will not be possible without new reliable radiative corrections for
this process.

\section{Summary}

Over a decade after the last precision measurements of the rare pion
and muon decays were completed, the PIBETA experiment at PSI is
revisiting the field, and producing new results.  Unlike any
experiment exploring the field before it, the PIBETA project measures
simultaneously practically all of the rare $\pi$ and $\mu$ decays.
This gives it an unusually powerful set of built-in consistency
checks.

The pion beta decay precision has been improved six-fold, resulting in
a first definitive test of the CVC and radiative corrections in a
meson.  Work on this decay channel is continuing, and will reach $\sim
0.5\,$\% accuracy in this first phase of the project.  The current
result, with a combined systematic and statistical uncertainty of
0.64\,\% is in excellent agreement with the SM predictions.

The PIBETA radiative pion decay measurements cover a broader region of
phase space than previous experiments, with more than an order of
magnitude higher statistics, and have brought about a four-fold
improvement in the precision of the pion axial form facto.  However,
the unexpected and pronounced deviation from the V$-$A description of
the process is limiting this precision, and raising important
questions.  This matter is being addressed in a current dedicated run.

Comparable improvements in precision are expected in radiative muon
decay.

Finally, in the following, second phase of the project, the PIBETA
collaboration will turn its attention to the $\pi \to e\nu$ process
which provides the best test of lepton universality and a selective
check on possible new physics.

This material is based upon work supported by the National Science
Foundation under Grants No.\ 0098758 and 0354808.

\end{document}